\documentclass{ws-mpla}
\usepackage[super]{cite}
\usepackage{graphicx}
\usepackage{xcolor}
\usepackage{float}
\usepackage{subfigure}
\usepackage[nodisplayskipstretch]{setspace}
\begin{document}

\markboth{SDG, A. Ranjan, H. Nandan, V. Sharma}{Heavy Hexaquarks in the Flux Tube Model}

\catchline{}{}{}{}{}

\title{Heavy Hexaquarks in the Flux Tube Model
}
\author{Sindhu D G$^{1*}$, Akhilesh Ranjan$^{1\dagger}$, Hemwati Nandan$^{2,3,4\ddagger}$ and V. Sharma$^{2\bullet}$}
\address{$^{1}$Department of Physics, Manipal Institute of Technology, Manipal 
Academy of Higher Education, Manipal, Karnataka 576104 India.\\
$^{2}$Department of Physics, Hemwati Nandan Bahuguna Garhwal Central 
University, Srinagar Garhwal, Uttarkhand 246174, India.\\ 
$^{3}$Department of Physics, Gurukula Kangri (Deemed to be University), 
Haridwar, Uttarakhand 249404, India.\\ 
$^{4}$Center for Space Research, North-West University, Potcheftstroom 2520, South 
Africa.\\
$^{*}$sindhudgdarbe@gmail.com\\$^{\dagger}$ak.ranjan@manipal.edu\\$^{\ddagger}$hnandan@associates.iucaa.in\\$^{\bullet}$vsharma.phys@hnbgu.ac.in}
\maketitle

\pub{Received (Day Month Year)}{Revised (Day Month Year)}

\begin{abstract}
Hexaquarks are one of the currently emerging topics in both experimental and 
theoretical high energy physics. Hexaquarks have been examined in relation to 
particle physics, however, there are still some research and theoretical 
conjectures surrounding their relationship to dark matter. Due to some 
experimental discoveries, it has attracted much interest and also resulted in 
new theoretical models to study the properties of these states. 
In the present work, Regge trajectories of some hexaquark states are compared 
with tetraquark and pentaquark states. The study is mainly concentrated on 
fully heavy hexaquark states. The mass spectra of these hexaquark states have 
also been investigated and the results are compared with other theoretical 
works. Our findings agree well with those of other researchers.

\end{abstract}

\ccode{PACS Nos.: 11.55.Jy. 12.40.Yx}

\section{Introduction}	
Since only color singlet multiquark states can exist in nature, one can also 
anticipate the possible existence of hexaquark states. This was initially 
proposed by Xuong and Dyson in 1964, which was called the dibaryon state.\cite{xuong} In the course of time, another hexaquark state with the quark 
content $uuddss$ was proposed by Jaffe.\cite{jaffe} Hexaquarks comprise 
either six quarks ($qqqqqq$) or three quarks and three antiquarks 
($qqq\bar{q}\bar{q}\bar{q}$). The six quark combination looks like two baryons 
bound together and can be called dibaryons. Hence dibaryons are the states with 
baryon number two. Though deuteron consists of six quarks $uuuddd$, it cannot 
be regarded as the compact hexaquark state, as the separation between the 
proton and neutron in the deuteron nucleus is large, of the order of $4fm$.	These 
hexaquark states have been searched extensively in N-N scattering 
experiments.\cite{bashkanov1,braun} Specially WASA-at-COSY collaboration 
reported a series of experimental results in this regard.\cite{adlarson1,adlarson2,adlarson3} The observation and confirmation of state 
$d^{*}(2380)$ indicated 
the existence of hexaquark and di-baryon states. \cite{bashkanov2,adlarson1,adlarson2,adlarson3,kren} The mass and angular momentum 
of $d^{*}(2380)$ are $2380 MeV$ and $J^{P}=3^{+}$ respectively. According to 
the most current update on the BESIII collaboration's search for hexaquarks, 
no hexaquark state has been discovered.\cite{ablikim}\\ 

Several research groups have theoretically investigated the properties of fully 
heavy tetraquark states,\cite{jing,zhi,chen1,bedolla,jin,tommy,yu} and the LHCb 
reported its experimental finding in 2020.\cite{aaij} Inspired by this, a lot 
of theoretical studies have been made in the field of fully heavy pentaquark 
and hexaquark states. \cite{yan,an,wang,gordillo} The quark-delocalization 
model,\cite{goldman1,goldman2} the flavor SU(3) skyrmion model,
\cite{kopeliovich} the chiral SU(3) quark model,\cite{zhang} the quark cluster 
model,\cite{fujita,yazaki} are some of the initial theoretical works, where the 
six quark states are predicted. Wang estimated the masses of fully heavy 
hexaquark states using the QCD sum rules. In this case, vector currents of the 
diquark-diquark-diquark type are built to examine the vector and scalar 
hexaquark states.\cite{wang} Using a diffusion Monte Carlo method, Pelegrina 
and Gordillo have defined the characteristics of fully-heavy compact hexaquarks.
 In the current investigation, they solely took into account compact hexaquark 
 objects and predicted the masses of states like $cccccc$, $cccccb$, $cccbbb$, 
 $ccbbbb$, and $bbbbbb$.\cite{gordillo} Using the constituent quark model, 
 Fang and colleagues have examined the mass spectra of fully heavy hexaquark 
 states. Spin-spin interactions, the linear confinement potential, and the 
 color Coulomb potential have all been taken into account. Their research 
 demonstrated the existence of hexaquark resonances. These resonances will 
 decay into heavy baryons.\cite{lu} The deuteron-like double charm hexaquark 
 states are investigated using the complex scaling method. Here authors have 
 considered the hexaquark states as molecules. The study mainly focussed on 
 determining the properties of bound and resonance hexaquark states.\cite{cheng1} 
The spin-zero hexaquark state has been searched using the LQCD 
techniques.\cite{loan} The quenched lattice QCD study shows that the proton-antiproton 
state cannot be regarded as a spin-zero hexaquark state. Amarasinghe and 
others have studied the scattering of two-nucleon systems using the variational 
approach.\cite{amarasinghe} With interpolating operators such as dibaryon 
operators, hexaquark operators are studied in this approach. This study does 
not give a proper conclusion regarding the existence of two-nucleon bound 
states. The $d^{*}(2380)$ is the most speculated hexaquark in connection with 
the dark matter. Hexaquarks as dark matter is still only a theoretical 
hypothesis that hasn't been substantiated with experimental confirmation. 
The main hypothesis is that hexaquarks could have been created in the early 
universe and accounted for the dark matter. Azizi and others have suggested 
a new particle termed the scalar hexaquark $uuddss$ as a potential dark matter 
candidate. The QCD sum rule approach is used to determine the hexaquark 
particle's mass and coupling constant. Depending on the strange quark mass 
employed, their calculations for the hexaquark mass range from 1180 $MeV$ to 
1239 $MeV$. \cite{azizi} Bashkanov and Watts, suggested that Bose-Einstein 
condensates of the $d^{*}(2380)$ hexaquark could serve as a potential candidate 
for dark matter. 
Additionally, the paper explores 
potential astronomical signatures, such as cosmic ray 
events, that could certainly indicate the presence of $d^{*}(2380)$ condensates. 
\cite{watts}\\

The study of the behavior of elementary particles serves as the foundation for 
numerous scientific disciplines, and theoretical models like the flux tube 
model are crucial for improving our fundamental understanding of the cosmos. 
In the present work, Regge trajectories of various hexaquark states in the flux 
tube model are examined, and the findings are compared with those of prior 
theoretical studies.\\

 The structure of this paper is as follows. In section 2, the flux tube model 
 is used to generate equations for classical mass and angular momentum for 
 various hexaquark configurations. In section 3, the results are then 
 extensively explored. And the conclusions are drawn in accordance. This work 
 examines the impact of string length variation on a few hexaquark states and 
 compares the calculated masses of these states to other theoretical results. 
 Further tetra, penta, and hexaquark Regge trajectories have been compared, and 
 the findings are quite interesting.
\section{Formulation}
\begin{figure}[H]
\begin{center}
\includegraphics[scale=0.38]{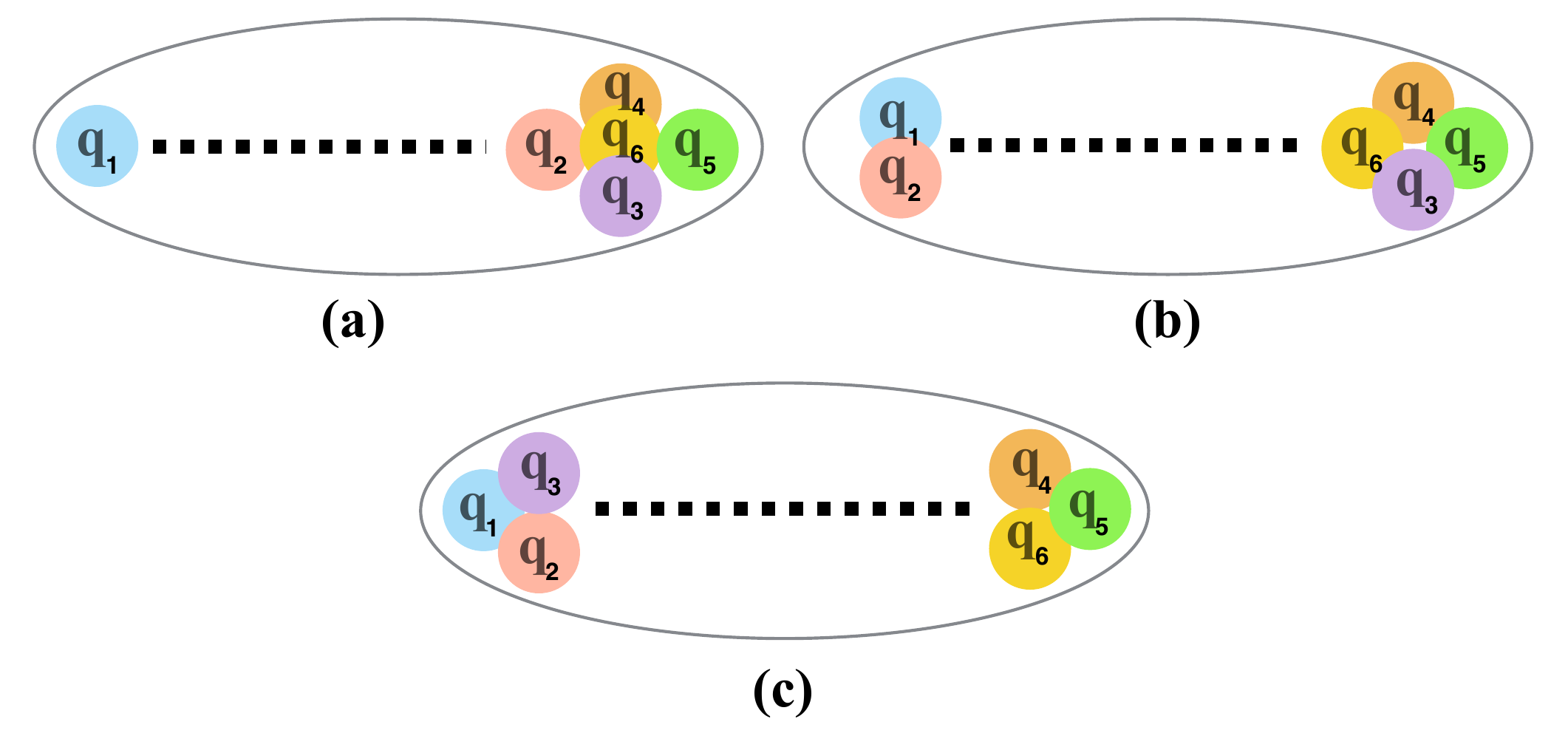}
\caption{Different configurations of hexaquarks with n (=1,2,3) number of 
quark at one end of the string. There exists several combinations (6, 15, and 
10) within those quarks for each of these three cases (a, b, and c 
respectively). }
\label{fig:1}
\end{center}
\end{figure}

The flux tube model plays an important role in explaining the  color confinement 
mechanism. The flux-tube model, in its most basic form, consists of a flux tube 
with quark and antiquark ends. It becomes a quantized Nambu string for light 
quarks and the potential model with linear confinement in the non-relativistic 
limit.\cite{olsson} In the flux tube model, it is assumed that the massless 
quarks are lying at the ends of the string and are considered to revolve with 
speed of light. 
If the string is rotating 
about its midpoint, then the classical mass 
and the angular momentum of a hadron is related by the following equation:
\begin{equation} 
J=\alpha_{0}+\alpha M^{2}
\label{appeqn1}
\end{equation}
where $\alpha_{0}$ and $\alpha$ are constants 
with $\alpha$=1/(2$\pi K$). Here $K$ is the string tension. These Regge 
trajectories have proven to be quite successful in providing a comprehensive 
framework for understanding the various properties of mesons, baryons, and 
other exotic hadrons.\cite{nohl,hothi1,hothi2,ghosh,chen2,juhi}
Our previous work utilized the flux tube model to study the Regge trajectories 
of tetra and pentaquarks. According to our findings, the flux tube model 
offers a good framework for exploring the properties of multiquark 
systems.\cite{ranjan} In this present work, we have extended this formulation 
to investigate hexaquark systems, with the goal of further understanding the 
confinement and behavior of these complex particles in the flux tube. We have 
found some new and interesting results. In the flux tube model, for hexaquarks, there will be thirty one different configurations with 
different combinations of quarks as shown in Fig. \ref{fig:1}. The 
Fig. \ref{fig:1} shows the set of configurations of hexaquarks with 
$n (= 1,2,3)$ number of quark at one end of the string. The expression representing mass and angular momentum for all configurations in Fig. \ref{fig:1} can be modified to the following general form:\cite{nandan1,nandan2}
\begin{equation}
\begin{aligned}
M_{n(mod)}&=\sum
\frac{Kl\mathcal M_n}{fM}\left(\int_{0}^{f}\frac{dv}{\sqrt{1-v^{2}}}+\int_{0}^{\frac{\sum m_{q_{n}}}{{\mathcal M_n}}f}\frac{dv}{\sqrt{1-v^{2}}}\right)\\
&+\gamma_{0}\sum m_{q_{n}}+\gamma_{n}\mathcal{M}_{n}.
\end{aligned} 
\label{appeqn2}
\end{equation}
\setstretch{1}
\begin{equation}
\begin{aligned}
J_{n(mod)}&=\frac{kl^{2}}{f
^{2}}\cdot \left(\frac{\mathcal M_n}{M}\right)^{2} \left\lbrace \int_{0}^{f}\frac{v^{2}dv}{\sqrt{1-v^{2}}}+\int_{0}^{\frac{\sum m_{q_{n}}}{\mathcal M_n}f}\frac{v^{2}dv}{\sqrt{1-v^{2}}}\right\rbrace\\
& +\frac{m_{q_{1}}fl}{M}\left\lbrace \gamma_{0}\mathcal{M}_{n}+\gamma_{n}\sum m_{q_{n}}\right\rbrace.
\end{aligned} 
\label{appeqn3}
\end{equation}
Here, $M=\sum_{i=1}^{6} m_{q_{i}}$, $\mathcal M_{n}$ = $M$ - $\sum m_{q_{n}}$, $\sum m_{q_{n}}=\sum_{n=1}^{n} m_{q_{n}}$where n is the number of quarks at one side of string.  Whereas, $\gamma_{0}=\frac{1}{\sqrt{1-f^{2}}}$, $\gamma_{n}=\frac{1}{\sqrt{1-\left(\frac{\sum m_{q_{n}}}{\mathcal M_n}\right)^{2}}}$, and
$f$ is the fractional rotational speed (actual speed is $fc$ with $c$ (speed of 
light in vacuum)=1 in natural system of units).\\
Equations (2) and (3) are shown here after integration:
\begin{equation}
\begin{aligned}
M_{n}=
\frac{Kl\mathcal M_n}{fM}\left\lbrace\sin^{-1}f+\sin^{-1}\left({\frac{\sum m_{q_{n}}}{\mathcal M_n}f}\right)\right\rbrace+\gamma_{0}\sum m_{q_{n}}+\gamma_{n}\mathcal{M}_{n}.
\end{aligned} 
\label{appeqn4}
\end{equation}

\begin{equation}
\begin{aligned}
J_{n}&=\frac{kl^{2}}{f
^{2}}\cdot \left(\frac{\mathcal M_n}{M}\right)^{2} \left\lbrace \frac{1}{2}\sin^{-1}f-\frac{f}{2}\sqrt{1-f^{2}}+
\frac{1}{2}\sin^{-1}\left({\frac{\sum m_{q_{n}}}{\mathcal M_n}f}\right)\right.\\
&\left.- \frac{1}{2} {\frac{\sum m_{q_{n}}}{\mathcal M_n}f} 
\times\sqrt{1-\left({\frac{\sum m_{q_{n}}}{\mathcal M_n}f}\right)^{2}}\right\rbrace +\frac{{\sum m_{q_{n}}}fl}{M}\left\lbrace \gamma_{0}\mathcal{M}_{n}+\gamma_{n}\sum m_{q_{n}}\right\rbrace .
\end{aligned} 
\label{appeqn5}
\end{equation}

We assume that all the thirty-one hexaquark configurations have equal 
probability to occur, therefore, the actual mass and angular momentum must be 
averaged. As $\sin\theta \le 1$ $\implies f\le \frac{M-m_{q_{1}}}{m_{q_{1}}}$ 
(corresponding to n = 1). From the special theory of relativity, 
$f\le1$. It is necessary to satisfy these conditions.
\section{Results and discussion} 
\begin{figure}[H]
\begin{center}
\includegraphics[scale=.35]{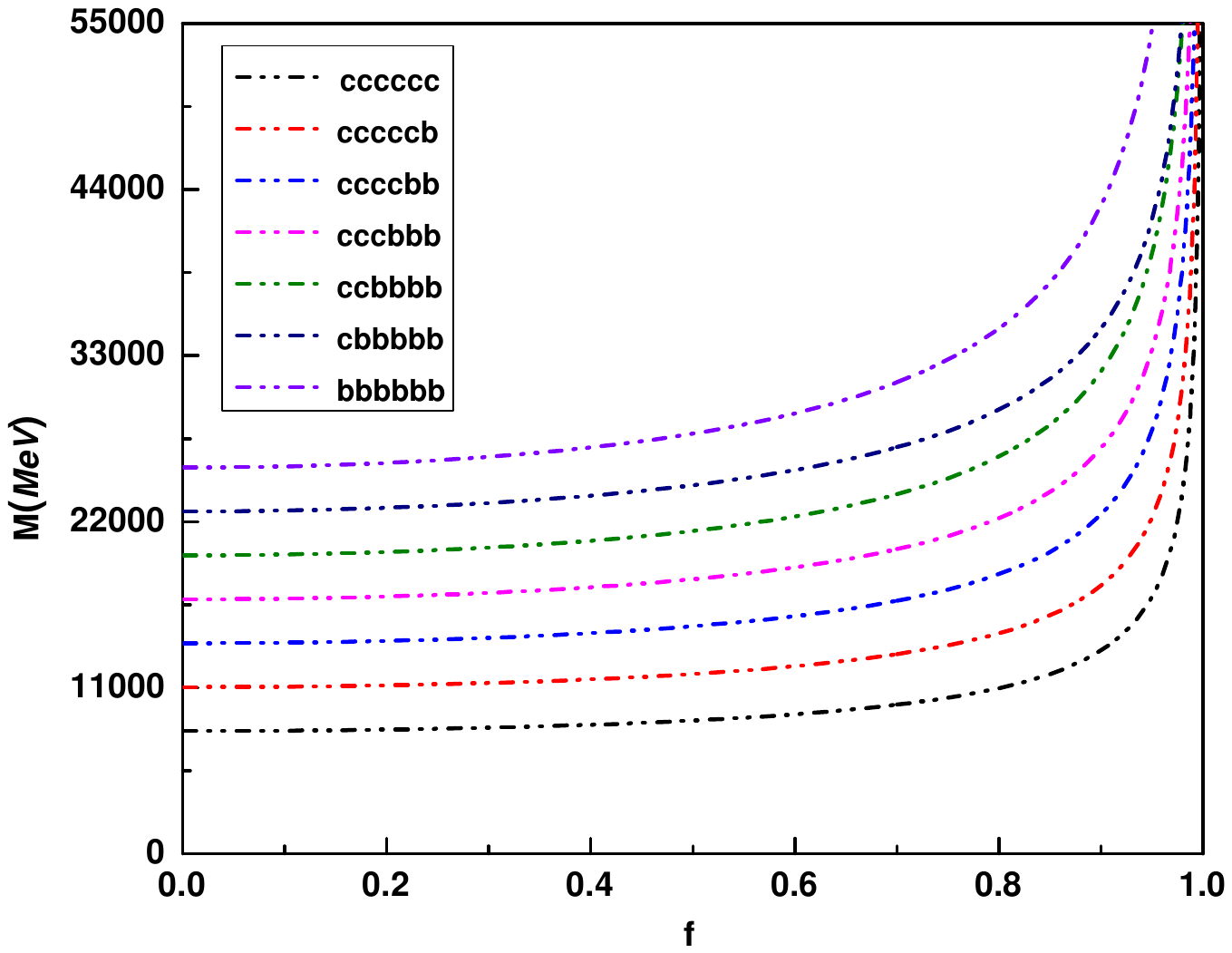}
\end{center}
\caption{Mass variation (M) with change in speed $f$ of the hexaquarks.}
\label{fig:M_f}
\end{figure}

\begin{figure}[H]
\begin{center}
\includegraphics[scale=.35]{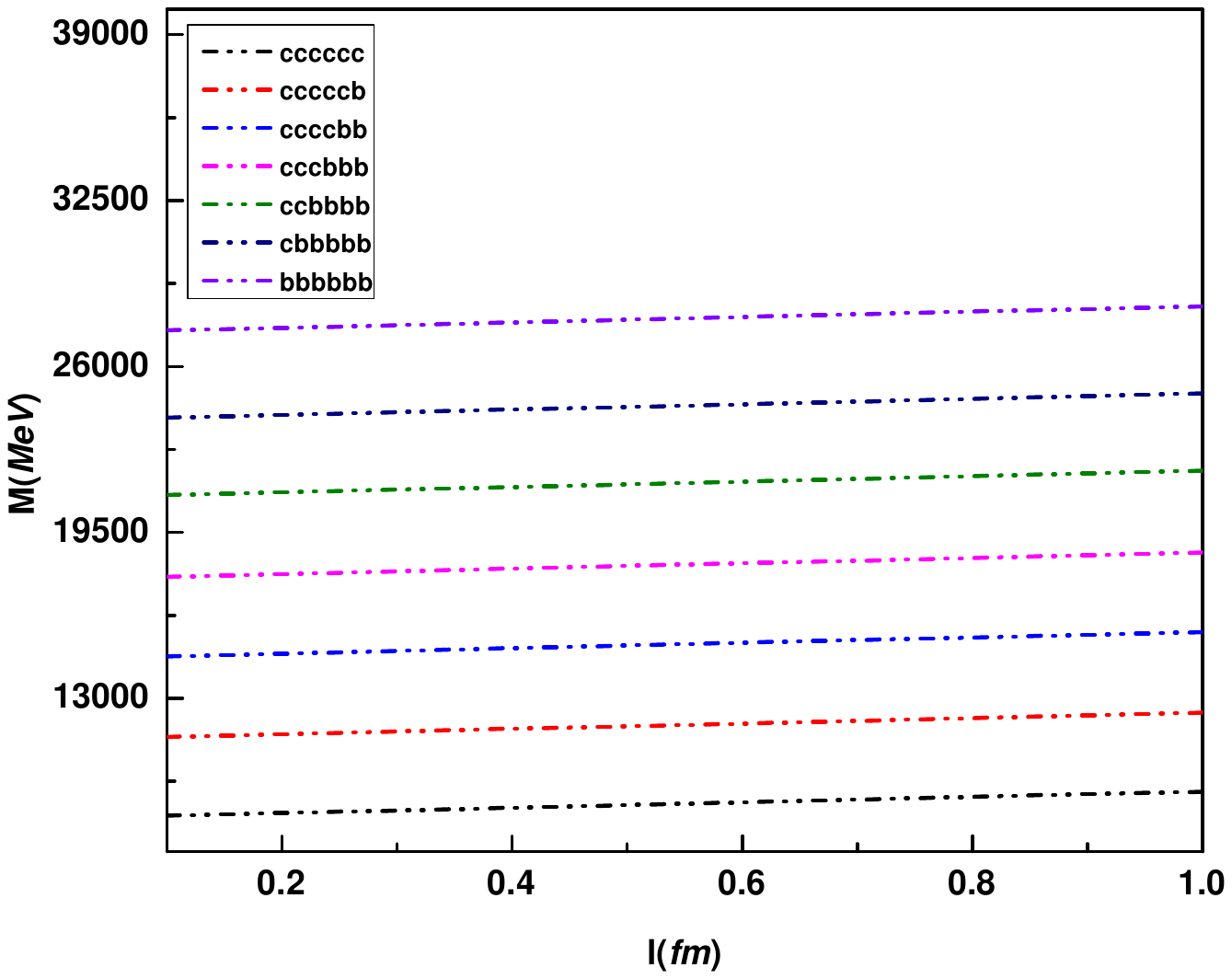}
\end{center}
\caption{Mass variation (M) with change in string length $l$ of the hexaquarks.}
\label{fig:M_l}
\end{figure}

\begin{figure}
    \centering
    \subfigure[]{\includegraphics[width=0.49\textwidth]{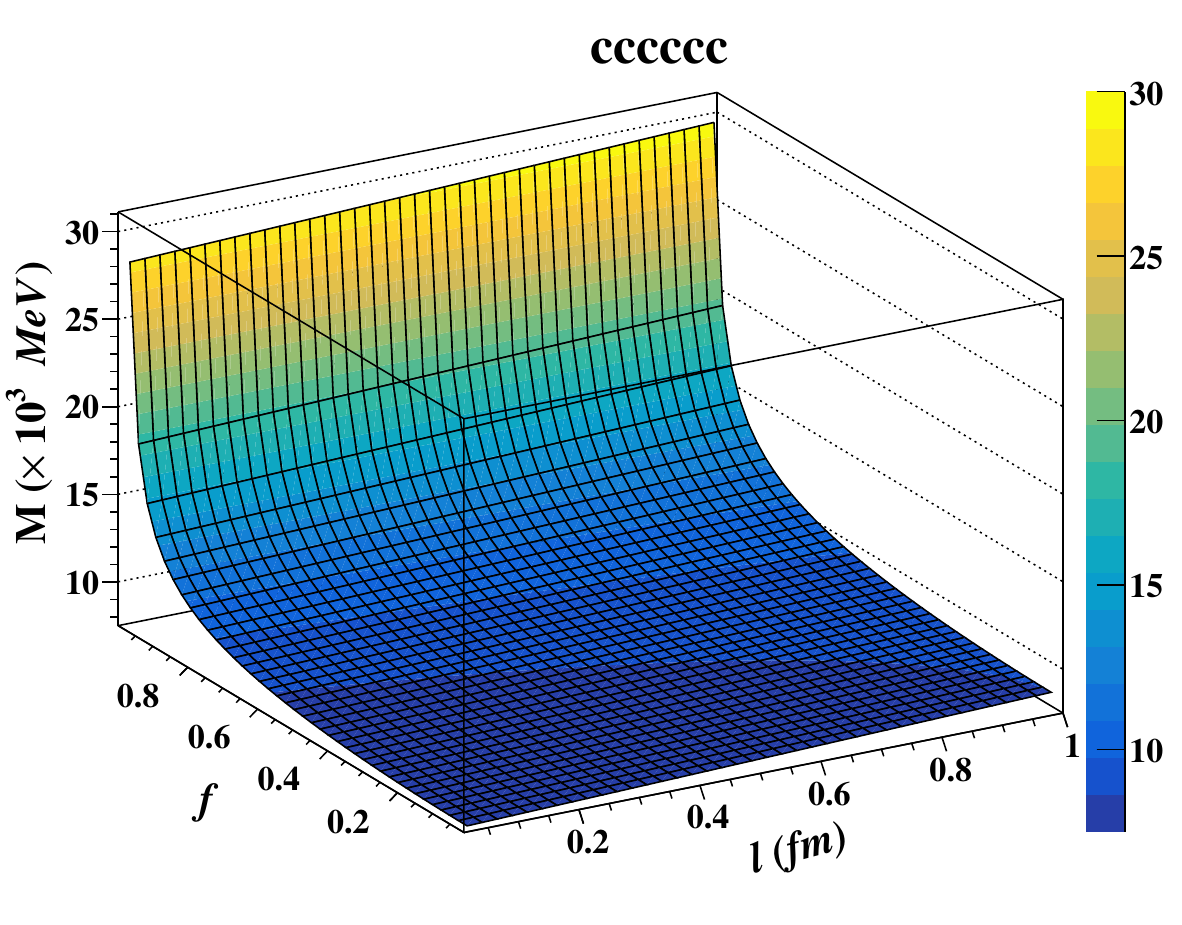}} 
        \subfigure[]{\includegraphics[width=0.49\textwidth]{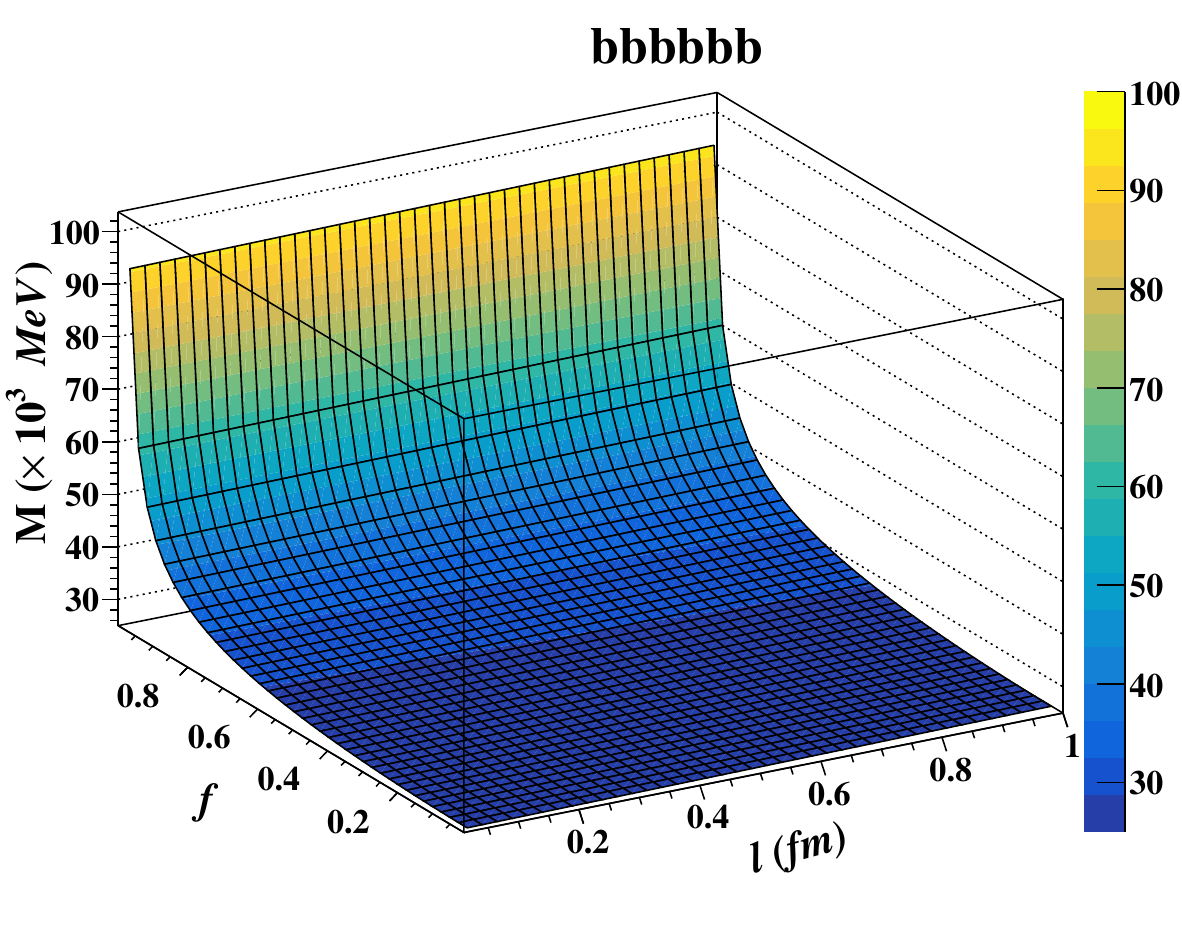}} 
  \caption{Mass variation with change in string length and fractional 
rotational speed for (a) $cccccc$, and (b) $bbbbbb$ hexaquark averaged for all 
possible (6+15+10) configurations of Fig. \ref{fig:1}. Other hexaquarks also 
follows the similar pattern of gradual change in mass with change in string 
length.}
  \label{fig:M_l_f}
\end{figure}

\begin{figure}[H]
\begin{center}
\includegraphics[scale=.35]{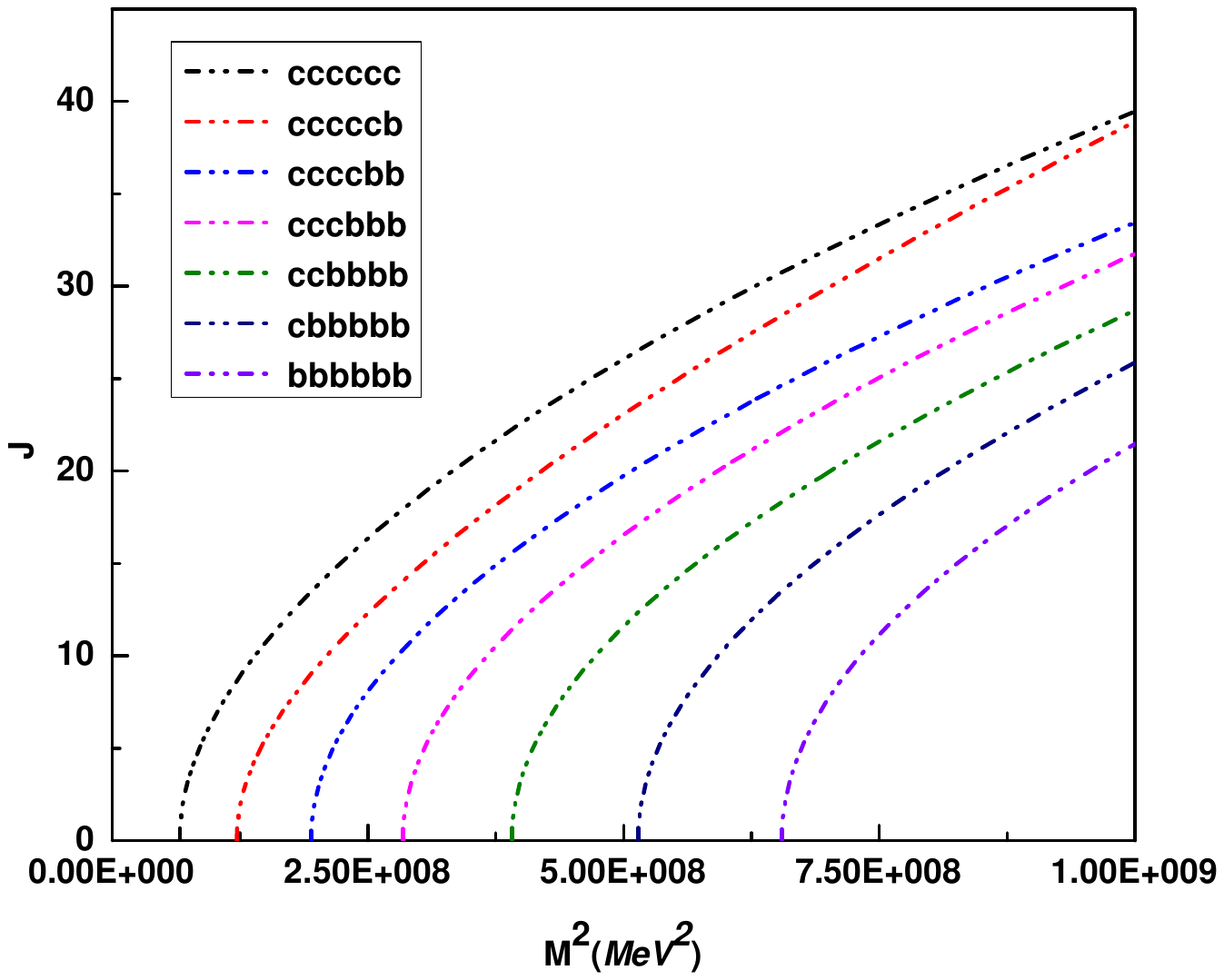}
\end{center}
\caption{Regge trajectories for different hexaquarks.}
\label{fig:reg_tr}
\end{figure}

\begin{figure}[H]
\begin{center}
\includegraphics[scale=.35]{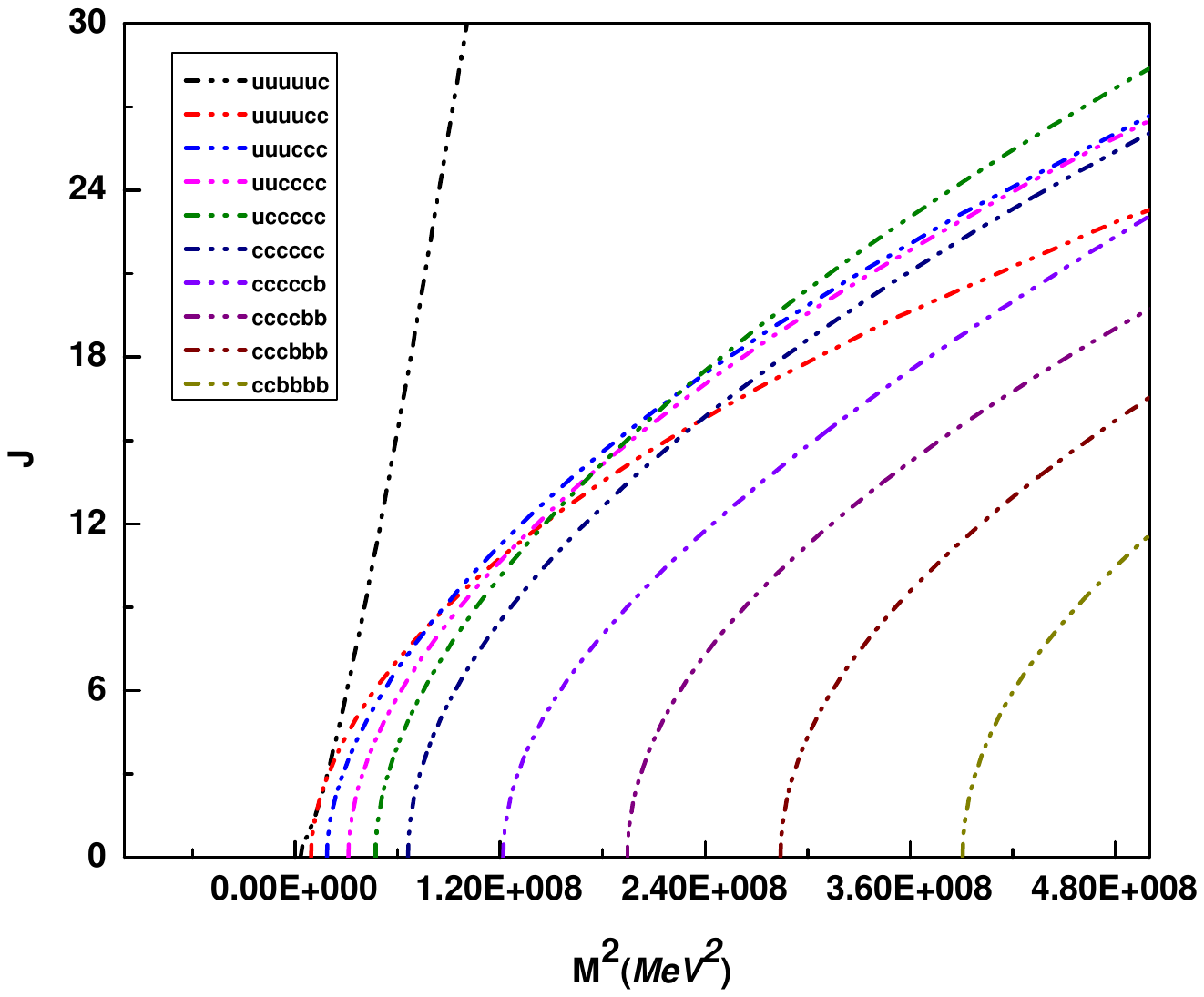}
\end{center}
\caption{Comparison of Regge trajectories for light to heavy hexaquarks.}
\label{fig:LightHeavyQ}
\end{figure}

\begin{figure}[H]
\begin{center}
\includegraphics[scale=.35]{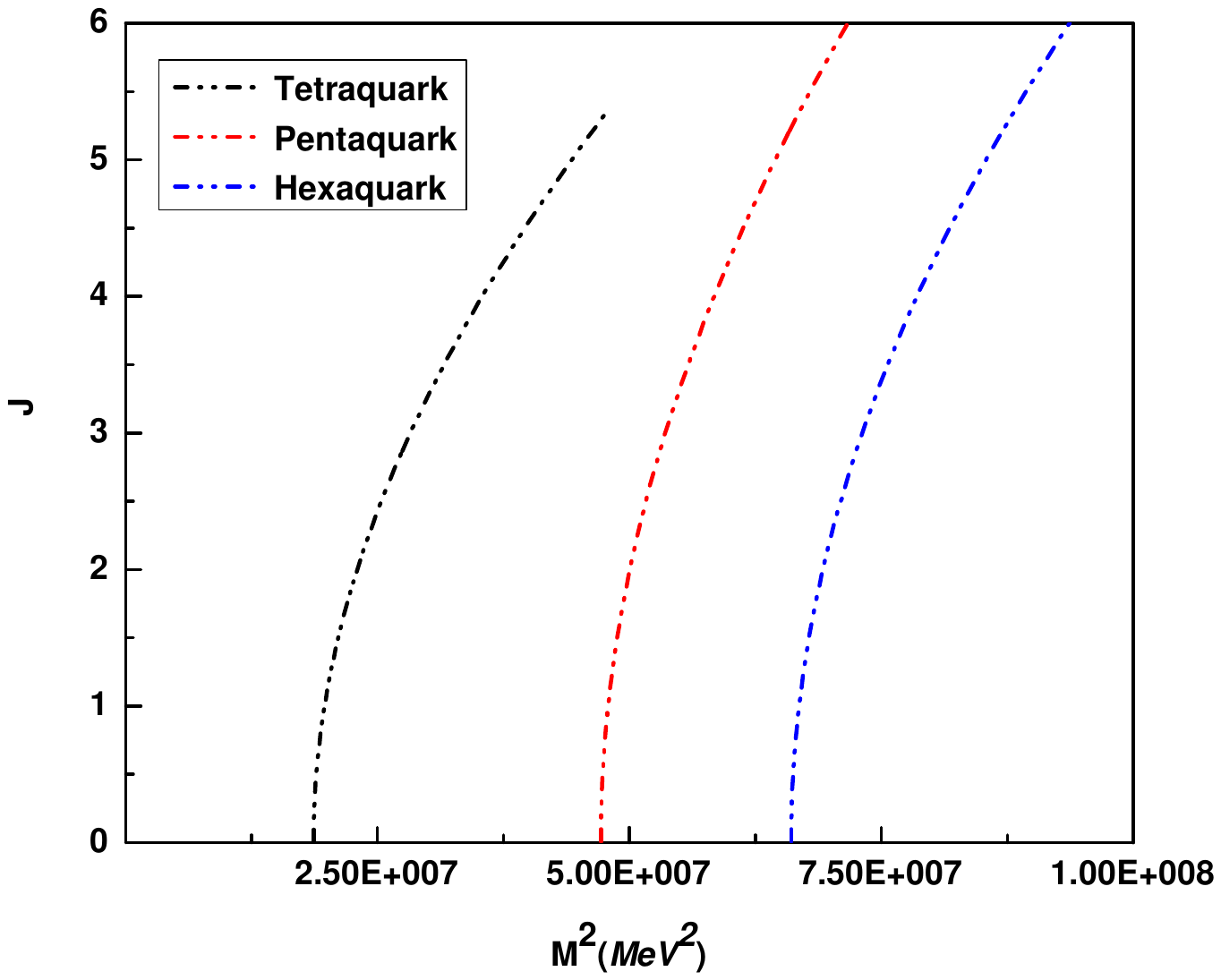}
\end{center}
\caption{Regge trajectories for tetra, penta, and hexaquarks.}
\label{fig:reg_tr3}
\end{figure}

The masses of quarks taken for the calculation are, $m_{up}=2.16MeV$, 
$m_{down}=4.67MeV$, $m_{strange}=93MeV$, $m_{charm}=1270MeV$ and $m_{bottom}=4180MeV$ respectively
, and $K=0.2GeV^{2}$.\cite{pdg} Here, $fc$ is the rotational speed of the low 
mass end of the string and $fc<1$. Here, the quarks and antiquarks that make 
up $q_{1}$ $q_{2}$, $q_{3}$ $q_{4}$, $q_{5}$, and $q_{6}$ are not stated 
explicitly. It is a general approach and one can consider any one as a quark 
or an antiquark. It will not have an impact on the formulae used here. 
In Table (\ref{tab:1}), calculated mass of different fully heavy hexaquark 
states are 
compared with other theoretical results.  The mass of a hexaquark increases 
with decrease in the string length $l$ indicating the fact that, the QCD 
effect will be more for higher mass states. The speed also decreases for 
heavier states. There are few hexaquark states having same quark flavor. For 
states with same quark flavor, it is not necessary to consider the thirty one 
configurations as mentioned earlier. In Table (\ref{tab:2}) the calculated 
mass of hexaquark states having atleast one heavy quark are mentioned. We have 
considered different $l$ values for different states. We have found that the 
present results are in good agreement with other models. The trend in the 
results is similar to that obtained for the fully heavy hexaquark states. One 
of the noteworthy observation is that, the string length as well as rotational 
speed grows as we move to the state with a higher angular momentum. It is 
evident from the Regge trajectory equation that the angular momentum of a 
particle is proportional to the length of the string, while the mass squared 
is proportional to the tension of the string. Hence, as we are moving to the 
higher $J$ state, $l$ increases. Again, if the mass and distance from the 
center of rotation remain constant, as the speed rises, its orbital angular 
momentum will rise. This is due to the fact that orbital angular momentum is 
described by the equation $L = mvr$, where $m$ is the object's mass, $v$ is 
its speed, and $r$ is its distance from the center of rotation. It is crucial 
to remember that the mass and distance from the center of rotation both have 
an impact on the relationship between orbital angular momentum and speed. For 
instance, the orbital angular momentum will increase if the object's mass rises 
while its speed stays the same. It is clearly visible from the results given in 
both the tables that, for hexaquarks with heavy quark flavors, string length is 
very small. The length of the string for the heavy quark is expected to be 
shorter than that of a light quark, due to the relationship between quark mass, 
string tension, and string length in string theory. The intensity of the strong 
nuclear force, which holds quarks together, is correlated with the tension of 
the string, a fundamental constant of nature. A quark's mass is determined by 
the energy stored in the string, which is related to the length and tension of 
the string. The energy held in the string increases as the mass of the heavy 
quark increases, in order to keep the same tension, the string's length must be 
reduced.  
Fig. (\ref{fig:M_f}) shows the mass variation of hexaquarks with 
variation in speed $f$. It is almost same for all the hexaquark states and is 
highly non-linear. With increasing string length, the hexaquark mass variation 
rises linearly (Fig. \ref{fig:M_l}).  Variation of mass of hexaquark with speed $f$ and string length $l$ 
averaged for all possible configurations is depicted in Fig. \ref{fig:M_l_f}.
The calculated expression (Eq. \ref{appeqn3}; n=1) clearly demonstrates that 
the mass rises as the string length rises.  
Fig. (\ref{fig:reg_tr}) shows the Regge trajectories of different hexaquark 
states. It is found to be nonlinear. There are thirty-one hexaquark 
configurations in the flux tube model, and it is assumed that all the 
thirty-one configurations are equally probable. Hence the mass and angular 
momentum is averaged and the nonlinearity is obvious.
 Fig. (\ref{fig:LightHeavyQ}) represents the Regge trajectories of hexaquarks 
 made up of both light and heavy quarks. It demonstrates clearly how 
 nonlinearity rises with heavier states. Fig. (\ref{fig:reg_tr3}) shows the 
 Regge trajectories of tetra, penta, and hexaquark states with all charm quark 
 configurations. It is apparent from the figure that the Regge trajectories of 
 tetraquark, pentaquark, and hexaquark are showing the same behavior.
\begin{table}[H]
\tbl{Predicted masses of different hexaquark states}
{\begin{tabular}{|c|c|c|c|c|c|c|}
\hline \noalign{\smallskip} 
SI. no.&Quark&$J$&M$_{cal}$&$l$&$f$&Other\\
&structure&&($MeV$)&($fm$)&&Results\\
&&&&&&(See Ref.\cite{lu,wang,gordillo})\\
&&&&&&($MeV$)\\
\noalign{\smallskip}
\hline
\hline
1&$cccccc$&1&9526.39&0.11&0.709&$9490\pm 130$\\
\hline
2&$cccccb$&1&13214.81&0.08&0.740&13176\\
\hline
3&$ccccbb$&1&16352.36&0.07&0.704&16373\\
\hline
4&$cccbbb$&1&19119.56&0.065&0.657&19221\\
\hline
5&$ccbbbb$&1&22029.22&0.06&0.613&22775\\
\hline
6&$cbbbbb$&1&25106.67&0.052&0.612&25980\\
\hline
7&$bbbbbb$&1&28090.95&0.049&0.560&$28500\pm 150$\\
\hline
\end{tabular}
\label{tab:1}} 
\end{table}
\begin{table}[H]
\tbl{Predicted masses of different hexaquark states}
{\begin{tabular}{|c|c|c|c|c|c|c|}
\hline \noalign{\smallskip} 
SI. no.&Quark&$J$&M$_{cal}$&$l$&$f$&Other\\
&structure&&($MeV$)&($fm$)&&Results\\
&&&&&&(See Ref.\cite{gerasyuta1,gerasyuta2})\\
&&&&&&($MeV$)\\
\noalign{\smallskip}
\hline
\hline
1&$uuuuuc$&1&3293.18&0.20&0.980&$\cdots$\\
\hline
2&$uuuudc$&1&3388.39&0.3&0.979&$\cdots$\\
&&2&3858.41&0.2&0.987&3902\\
\hline
3&$uuuddc$&1&3396.33&0.3&0.979&$\cdots$\\
&&2&3868.37&0.2&0.989&3863\\
\hline
4&$uuuucc$&1&4516.49&0.21&0.890&$\cdots$\\
&&2&5125.59&0.30&0.919&5250\\
\hline
5&$uuudcc$&1&4401&0.22&0.880&4420\\
&&2&4950.40&0.32&0.909&4911\\
\hline
6&$uuddcc$&1&4304.43&0.23&0.870&4364\\
&&2&5032.48&0.31&0.914&5086\\
\hline
7&$uusscc$&1&4814.71&0.19&0.903&$\cdots$\\
\hline
8&$uscdsc$&1&4841.12&0.18&0.907&$\cdots$\\
\hline
9&$uuuccc$&1&5892.87&0.16&0.887&$\cdots$\\
\hline
10&$uuuudb$&1&5996.24&0.9&0.768&5988\\
&&2&6926.28&0.88&0.891&6926\\
\hline
11&$uuuddb$&1&6161.97&0.8&0.820&6142\\
&&2&6949.81&0.85&0.895&6928\\
\hline
12&$uucccc$&1&7014.48&0.14&0.795&$\cdots$\\
\hline
13&$uccccc$&1&8049.02&0.13&0.766&$\cdots$\\
\hline
14&$uuuubb$&1&10202.13&0.14&0.675&10290\\
&&2&11628.24&0.17&0.795&11620\\
\hline
15&$uuudbb$&1&11052.72&0.11&0.761&11395\\
&&2&11357.24&0.18&0.777&11372\\
\hline
16&$uuddbb$&1&10707.07&0.12&0.731&10828\\
&&2&11359.88&0.18&0.777&11518\\
\hline
\end{tabular}
\label{tab:2}} 
\end{table}

\section{Conclusions and Future Prospects}
In view of the results obtained, it is obvious to mark that the 
heavier states are accommodated with the string having shorter length. The 
masses of the hexaquark states are roughly linear when the rotational speed 
is low, but as the speed increases, they become significantly nonlinear. The 
Regge trajectories are largely linear for hexaquark states with the light 
quarks, but they become highly nonlinear for heavier states as evident from Fig. \ref{fig:LightHeavyQ}. We observed that the Regge trajectories for fully heavy tetraquark, pentaquark, and hexaquark states shows similar pattern (See Fig. \ref{fig:reg_tr3}). In order to fully comprehend the behavior and characteristics of the Regge trajectories of hexaquark states, additional theoretical and experimental research will be required for deeper insights. 
The detailed study of effects of string length on the hexaquark mass, breaking of flux tube, their relation with quark confinement, and emerging possibilities of dark matter sector are the topics beyond the scope of this work and we would like to address these issues in our  future investigations.
\section*{Acknowledgments}
AK is thankful to Manipal Academy of Higher Education (MAHE) Manipal for the 
financial support under scheme of intramural project grant no. 
MAHE/CDS/PHD/IMF/2019. SDG is thankful to  `Dr. T. M. A. Pai Scholarship 
Program' for the financial support.

\end{document}